\documentclass[%
onecolumn,%
oneside,%
floats,%
aps,%
 prd,%
nobibnotes,%
nofootinbib,%
amsmath,%
amssymb,%
amsfonts,%
amscd,%
  superscriptaddress,%
]{revtex4}

\usepackage[utf8]{inputenc}
\usepackage{graphicx,array,dcolumn}
\usepackage{cases}
\usepackage[paperwidth=210mm,paperheight=297mm,centering,hmargin=1.8cm,vmargin=2.5cm]{geometry}
\usepackage{enumerate}

\usepackage{hyperref}
\hypersetup{hidelinks}

\usepackage{geometry}                		
\usepackage{graphicx}				
\usepackage{amssymb}

\newcommand{\rar}{\rightarrow}

\def\be{\begin{eqnarray}}
\def\ee{\end{eqnarray}}

\begin{document}

\title{New Old Mechanism of Dark Matter Burning
}

\author{A.D. Dolgov}
\email{dolgov@fe.infn.it}
\affiliation{Novosibirsk State University, Novosibirsk, 630090, Russia}
\affiliation{ITEP, Bol. Cheremushkinsaya ul., 25, 117259 Moscow, Russia}

\begin{abstract}

Recently several papers have been published, where  "a new mechanism" of dark matter burning and freezing
is suggested. The usual two-body annihilation process is generalized to multi-body initial states, mostly three-body.
These processes have quite a few interesting cosmological implications.

I want to indicate here that such a process was studied in 1980 to determine the cosmological number density 
of the so called theta-particles. The file of the translated into English original paper and the reference to the 
published English version are attached.

\end{abstract}

\maketitle

In 1980 L.B. Okun~\cite{LBO-theta} suggested a model of a new interaction which has macroscopically 
(and possibly even astronomically) large confinement radius. In this model new light, but massive scalar
particles, the so called theta-particles, must exist, which possibly do not interact with the standard model 
world (or extremely weakly  coupled to it). Such particles cannot annihilate into the particles of the standard 
model, and their cosmological number density might be dangerously high. However, an unusual channel of 
the annihilation is open for them~\cite{AD-theta}:
\be
2\theta \rar 3 \theta,
\label{theta-ann}
\ee
which diminishes their cosmological density down to a reasonably small number,
see eq. (8) in the attached file of ref~\cite{AD-theta}. 

It worth noting that the relative frozen number density of dark matter determined 
through the usual two-body annihilation is inversely proportional to the 
Planck mass, $r_f^{(2\rar 2)} \sim 1/m_{Pl}$, while the freezing of the three-body process 
it is noticeably less efficient, resulting in the density inversely  
proportional to the square root of $m_{Pl}$, i.e. $ r_f^{(3 \rar 2)} \sim 1/\sqrt{m_{Pl}}$.

\begin{figure}[!h]
\center
\begin{tabular}{c c c}
\includegraphics[scale=1]{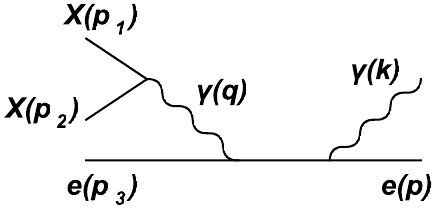} & &
\includegraphics[scale=1]{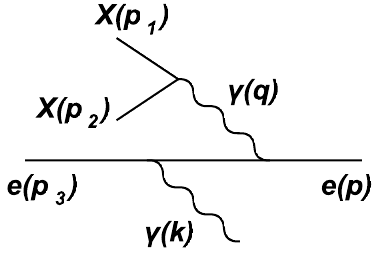}
\end{tabular}
\caption{\label{fig:1} Annihilation of $X \bar X $ into one photon catalyzed by electron.}
\end{figure}

Another type of $3 \rar 2$ process is studied in our work with A.S. Rudenko~\cite{ad-ar-3-2} in connection with 
the calculations of the cosmological density of millicharged particles, MCP or $X$. The recent review of the properties 
of MCPs and the observational bounds on their charge and mass can be found in ref.~\cite{ad-ar-MCP}. 
If these  X-particles are lighter than electrons, $ m_X < m_e$, they cannot annihilate into $e^+ e^-$-pairs, 
but only into two photons. The cross-section of the process $ X \bar X \rar 2 \gamma$ is proportional to the tiny factor
$\epsilon^4$, where $\epsilon$ is the ratio of the electric charges $\epsilon = q_{X}/q_e \ll 1$. In the limit of sufficiently
small $\epsilon$ the process $X\bar X \rar 2\gamma$ would be subdominant with respect to $X \bar X e \rar e \gamma$. 
It looks as the electron catalysis of $X \bar X$-annihilation, see fig.~\ref{fig:1}.

Recently the mechanism of multi-body annihilation was rediscovered in 
three almost simultaneous papers~\cite{1,2,3}. In ref.~\cite{1} such processes got a nice name "assisted annihilation".
These papers opened a new domain for possible dark matter particles but I would like to indicate that the pioneering 
idea was put forward almost 30 years ago. The English version of the original paper~\cite{AD-theta} 
and the scan of the first page of the translated journal are presented at the end of this file.
\\[3mm]
I thank A.I. Vainshtein who send me the pdf file of my work~\cite{AD-theta} published in English. 


\section{Appendix.  English translation of the paper  \cite{AD-theta}}   

\begin{center}
\Large{{\bf The concentration of residual $\theta$ particles}} \\[3mm]
A.D. Dolgov \\[1mm]
{\small{\it Institute of Theoretical and Experimental Physics, State Atomic Energy Commision}\\
Subitted 15 February 1980)\\
Yad. Fiz. {\bf 31} 1522 - 1528 (June 1980)}
\end{center}
An expression is derived for the concentration of residual particles in the Universe if their extinction 
occurs in three-particle collisions. This result is used to set limits on the parameters of $\theta$
particles (hypothetical with a large confinement radius). \\[2mm]
PACS numbers: 98.70.Vc,95.30.Cq,12.40.Bb \\[4mm]

From the great explosion at the "creation of the world" we have inherited all of the types of stable
elementary particles  (those already known and those  as yet unknown) that exist in nature. Indeed, 
present-day cosmology cosmology asserts with assurance that in the  distant past the temperature and 
density of matter in the Universe were extremely high and also, which is very that the primeval plasma
was in the state close to thermodynamic equilibrium. (A discussion of these matters can be found in 
Ref. 1 or Ref. 2.)  At equilibrium all particles with $m_i < T$ exist in the plasma with about the same
density $n_i \approx 0.2 T^2$, and thereafter the residual remnants of these particles have survived 
down to our own time.  During the expansion of the Universe the temperature $T$ falls with the time $t$
according to the law
\be
T^2 t = \frac{45}{32 \pi^3} \, N^{-1/2} m_P \equiv C m_P , \hspace{10cm} (1)
\nonumber
\label{1}
\ee
where $m_P = G^{-1/2} \approx 10^{19} $ GeV is the Planck mass and $N$ is, roughly speaking, the
number of types of different elementary particles with $m_i <T$ present in the plasma; more exactly, the
contribution of each boson  to $N$ is $K_s/2$ and that of each fermion is $ 7K_s/16$, where $K_s$
is the number of the spin states of the particle. The quantity $N$ of course depends on the temperature.\\

When the temperature becomes smaller than the mass of the particles, their equilibrium concentrations 
begin to fall very rapidly; $ n_i \approx  const\,(m_i T)^(3/2)\times \exp (-m_i /T) $, if the chemical
potential is $\mu_i = 0$. Therefore, it there were still equilibrium at present , in a world symmetrical
in the $\mu_i$ there would exist practically no stable massive particles. However, along with the fall
of the concentrations the rate of burnout of the particles deceases and becomes  smaller than the 
rate of the expansion of the universe. The thermodynamical equilibrium is destroyed. After this time
the particle concentrations in a comoving volume element are approximately constant, the concentration
being frozen[3]: $r_i \equiv n_i /(0.2 T^3) \approx const$.\\

If the burnout of the particles of  a given type occurs owing to mutual annihilation into light particles in
tow-particle collisions, the time of the freezing-in is determine by the condition
\be
\dot r_{eq} / r_{eq} \approx \sigma v \cdot 0.2 T^3 r_{eq}, \hspace{10cm} (2)
\nonumber
\ee 
where $ r_{eq} = (m/T)^{3/2} \,\exp (-m/T)$ is the equilibrium relative concentration, $\sigma$ is the
annihilation cross section, and $v$ is the relative velocity of the  colliding particles. Since the decrease 
of the temperature with time follows a power law, we have $\dot T /T \approx t^{-1}$ and 
$\dot r_{eq} / r_{eq} \approx (m/T) t^{-1} $, where $t$ is the time measured from the initial singularity.\\

The equilibrium concentration $r_{eq}$ drops sharply as the temperature falls, so that in the 
condition (2) we can set  $m \approx T$ everywhere except in the exponential; we thus get  
for the frozen-in concentration of the residual particles the estimate$^1)$
\be
r_f \approx (\sigma v m^2 t )^{-1} = \frac{M}{m_P}\, (C v \sigma m^2)^{-1}\hspace{9cm} (3)
\nonumber
\ee
A more accurate treatment taking into account the remaining annihilation of particles [1] after
the destruction of thermodynamic equilibrium, and also the fact the $T_f < m$, gives the asymptotic
concentration
\be
r_f \approx \frac{10 m}{T_f} \,\frac{m}{m_P}\,\frac{N^{1/2}}{\sigma v m^2} \hspace{10cm} (4)
\nonumber
\ee
where $T_f$ is determined from the condition [2]
\be
\frac{m}{T_f} - \frac{1}{2}\,\ln \frac{m}{T_f} = 42 + \ln (\sigma v m^2) + \ln \frac{m_N}{m} - 
\frac{1}{2} \,\ln N .
\hspace{6cm} (5)
\nonumber
\ee
Here $m_N$ is the mass of the nucleon and the term 42 comes mainly from $\ln (m_P/m_N)$.
\\

Similar results were first obtained in Refs. 3 and 4, and have since been widely used to
calculate the residual concentration of various actually existing and hypothetical particles.\\

The purpose of the present note is to calculate the concentration of the residual particles
in the case in which their burning out is due to three particle collisions. Until recently no physical
examples were known in which three-particle collision are important.  In a recent paper by 
Okun' [5], however, the hypothesis was advanced that there may exist particles of new type
with a large (perhaps even macroscopic) confinement radius, which are called $\theta$  
particles. Among the $\theta$ particles there must be $\theta$ gluons, vector bosons which
are carriers the $\theta$-gauge interactions. Moreover, in the terminology of  of Ref. [5], there
must exist $\theta$ leptons i.e., particles that interact with the $\theta$ gluons and also have 
electroweak interactions. If besides this the particles possess color, i.e. interact with colored
gluons, they can be called $\theta$ quarks.\\

Just as in ordinary quantum chromodynamics there must exist white or singlet bound states of
gluons, gluonium, so also in the framework of the scheme of Ref. [5] there must be singlet bound
states of the $\theta$ gluons, thetonium. Thetonium almost completely lacks  any interaction with
ordinary matter and is practically stable. In what follows we mush distinguish two cases. If the
inverse of the confinement radius, $\Lambda_\theta$, of the $\theta$  interaction is smaller than
the present temperature of the (it is perhaps more correct to speak of a liquid) of the residual gluons
(i.e. , a few degrees Kelvin), then at the present time the $\theta$ gluons are in a free phase, and
their number is of the order of the number of photons, i.e. $\sim 10^3/$cm$^3$. Since the mass of
the thetonium is of the order of $\Lambda_\theta$, the total energy density of the $\theta$ gluons is
negligibly small in comparison with the total density of matter in the Universe, and in this case the
contribution of the $\theta$ gluons to the total energy density of the Universe is negligible. \\
 
If $\Lambda_\theta$ is larger than the present temperture, the $\theta$ gluons exist in the form of 
bound states, thetonum (we shall denote them in what follows by the letter $\theta$). If the number
of thetonium bund states ("atoms") were conserved in the interactions, their density in a comoving
volume would be constant and amount to $\sim 10^2/$cm$^3$. In this case the condition
$\rho_\theta <  \rho_{max} = 10^{-29} $ g/cm$^3 = 6 \cdot 10^{-3}$ MeB/cm$^3$ means that the
mass of the gluonium atom is subject to the limitation $m_\theta < 100$ MeV, in analogy with the
limitation on the neutrino mass [2,6]. However, thetonium can disappear in the reaction
 \be 
3 \theta \rar 2 \theta \hspace{10cm} (6)
\nonumber
\ee
and therefore its present concentration will be much smaller. As we have already pointed out, the
the concentration of the residual particles  is close to its concentration at the time when the rate of
expansion of the world became larger than the rate of the reaction in which they interact,  i.e., at
the time when the thermodynamic equilibrium is destroyed. For two-particle burnout the residual
concentration after the loss of equilibrium is further diminished by a factor $T_f/m$. According to
Eq. (5) this factor is of the order $10^{-2}$. Here we shall calculate the remaining burnout for the
case of three-particle collision, which, as should be  expected , turns our to be much smaller than
for two-particle collisions, with a corresponding increase of the residual concentration. \\

The freezing-in temperature for three-particle collision is given by the condiition
\be
\frac{m}{T_f}\,\frac{1}{t_f} = r_f^2 \,(0.2 T^3_f)^2 \Gamma(3 \rar 2) \hspace{10cm} (7) 
\nonumber
\ee
where $m$ is the mass of the colliding particles (in this case $m = m_\theta$, the mass of
thetonium), and $\Gamma (3 \rar 2) \equiv \Gamma = (2m)^{-2} \int | A(3\rar 2)|^2 d\tau_2$ 
is normalized to unit probability density of the transition per unit time. \\

If we assume as before that at the time of freezing-in $T \approx m_\theta$, we gat the
approximate estimate
\be
r_f \approx \left(\frac{m}{m_P}\right)^{1/2} [(0.2)^2 C \gamma m^5]^{-1/2}. \hspace{9cm} (8) 
\nonumber
\ee
We note that the result is proportional to $m_P^{-1/2}$, differing from $m_P^{-1}$ dependence
for two-particle collisions.\\

Equation (7) determines the freezing-in temperature; the result differs from that for two-particle
burning-out by a factor 1/2:
 \be 
 \frac{m}{T_f} - \ln \frac{m}{T_f} = \frac{1}{2} \left[ 42 + \ln (\Gamma m^5) - \ln \frac{m_N}{m} 
 - \frac{1}{2} \ln N \right].  \hspace{6cm} (9) 
\nonumber
\ee
After the destruction of thermodynamic equilibrium, when the temperature falls below $T_f$, the
processes inverse to those of Eq. (6), i.e. $2\theta \rar 3\theta$, are of little importance. In fact, for
$T<T_f$ the density of the particles per unit volume decrease rather slowly, and therefore the 
occurrence of processes $2 \theta \rar 3\theta$ continues. The kinetic equilibrium maintained by
$ 2\theta \rar 2\theta$ processes also persists for $T<T_f$, so that the energy distribution of the
thetonium atoms is still exponential, $\sim \exp (-E/T)$. Consequently the endothermal processes
$2\theta \rar 3\theta$ die out rapidly and can be neglected. Accordingly, the concentration of the
thetonium atoms after the destruction of equilibrium is governed by the equation:
\be
\dot r \approx r^3  (0.2 T^3)^2 \gamma \hspace{10cm} (10) 
\nonumber
\ee
with the initial condition $r(T=T_f) = r_f = (m/T_r)^{3/2} \times \exp (-m/T_f)$. Equation (10) can be
integrated easily, and gives for the asymptotic thetonium concentration the result
\be 
r_\infty = r_f \left( 8T_f/m\right)^{1/2} \hspace{10cm} (11) 
\nonumber
\ee
Determining $r_f$ from Eq. (7), we finally obtain 
\be
r_\infty \approx 100 \left(\frac{m}{m_P}\right)^{1/2} \left(\frac{m}{T_f}\right)^{2}
(\Gamma m^5)^{-5/2},    \hspace{7cm} (12) 
\nonumber
\ee
where $T_f$ is determined from the relation (9). As is shown in ref. [5], the interactkon of thetonium
atoms is strong at distances of the order of $\Lambda_\theta \approx m_\theta$, so that it is natral to
suppose that $\Gamma m^5  \approx 1$  and $ m/T_f \approx 20$. From this it follows that the thetonium
mass must be smaller than 20-30 MeV. Indeed the energy density of the residual thetonium at present 
is given by
\be
\rho_\infty = m\theta r_\infty \approx 10^{-5} m_N (m_\theta  / m_N )^{3/2} \kappa,
(500 /{\rm cm}^2).  \hspace{7cm} (13) 
\nonumber
\ee
where $\kappa$ allows for the difference between the present temperatures, and consequently the
densities, of residual photons and thetonium atoms; $n_\gamma \approx 500$/cm$^3$ and
$n_\theta = \kappa n_\gamma$. The value of $\kappa$ is of course not known; it depends upon the
ratio between the number of types of particles that finally annihilate into photons and heat up the 
photon gas, and the number of types that annihilate into $\theta$ gluons, and also on the rate of
exchange of energy between between the photon and thetonium gases. It is hard to expect $\kappa$
to be noticeably smaller than unity. Remembering that the energy density of any form of matter has to
be limited by the quantity $\rho_{max} = 6 \cdot 10^{-6} $ GeV/cm$^3$, we find that
\be
m_\theta < 6^{2/3} 10^{-2} m_N \approx 30\,{\rm MeV}  \hspace{10cm} (14) 
\nonumber
\ee
Accordingly the confinement radius $r_\theta$ of $\theta$ particles must be larger then 10 fm.
The specific relation between $m_\theta$ and $r_\theta$  depends on the mechanism of confinement
 and is still unknown. It is shown in ref. [7] that in quantum chromodynamics, with a confinement 
 radius $\lambda_\theta \approx ( 0.1$ GeV$^{-1}$, the gluonium mass is 1-2 GeV. If (quite without
justification) we apply this to the case of $\theta$ particles, then the restriction (14) implies that the
$\theta$ confinement radius is larger than  100 fm. \\

We also note that we can get some restriction on the masses of $\theta$ quarks and (or) $\theta$
leptons if the latter exists and possess a conserved (more exactly, almost conserved) charge 
$B_\theta$ or $L_\theta$, analogous to the baryon charge $B$ or the lepton charge $L$. The
point is that if the schemes for unifying the strong and electroweak interactions are valid, the
baryon charge is in general not conserved and its nonconservation is large at high energies,
$E \gtrsim 10^{15} $ GeV. Together with CP violation, the nonconservation of baryon charge  can
give the observed baryon asymmetry of the Universe, leading to a ratio 
$r_B = n_B/n_\gamma = 10^{-9 \pm 1}$ instead of $r_b \approx 10^{-18} $, as given by Eq. (4).
There is now an extensive literature on this question; it can be found in Ref. [2] for example. In
analogy with the baryon asymmetry there must arise an asymmetry of the same order  in the
charges $B_\theta$ or $L_\theta$, i.e., there must be as much $\theta$ matter as ordinary matter
contained in the world. From the condition $rho_\theta < \rho_{max}$ there follows a restriction
on the mass of the lightest $\theta$ quark or $\theta$ lepton:
\be
m <  10^{1\pm 1} m_N < 100\,{\rm GeV} \hspace{10cm} (15) 
\nonumber
\ee
We recall that this restriction is correct if the lightest $\theta$ quark or $\theta$ lepton has an
(almost) conserved charge $B_\theta$ or $L_\theta $ and consequently is (almost) stable. \\

As shown in Ref. [5],a concentration of $\theta $ quarks at a level $\sim 10^{-20}$ of that of
nucleons contradicts the results the results of searches of the for anomalous hydrogen, since 
$\theta$ quarks by definition [5] have the ordinary strong interaction with colored gluons and 
must form anomalous hadrons. Consequently it would seem that such $\theta$ quark must not
exist at all. There are still some remaining possibilities, however. First of all the hypothesis that 
an asymmetry in $B_\theta$ or $L_\theta$ exists is not inescapable. In a symmetrical world the
concentration of $\theta$ quarks must be much smaller. To calculate it we must not directly use 
Eq. (4), since the cross section  for $q\theta \bar q\theta $  annihilation is of the form
\be
  \sigma v = \alpha^2_c /T^2       \hspace{10cm}     (16)
\nonumber
\ee
where $\alpha_c$ is the coupling constant of quantum chromodynamics. In the derivation of Eq. (4),
on the other hand, it was assumed that $\sigma v$ is independent of the temperature. The 
expression (16) is valid as $T> \Lambda_c$, where $\Lambda_c \gtrsim 0.1$ GeV is the inverse of
the confinement radius in quantum chromodynamics.. For $T < \Lambda_c$ the quarks condense 
along with the usual quarks into colorless $\theta$ hadrons. The concentration of $\theta$ quarks at 
this time can be determined using the equation 
\be
\dot  r / r^2 = 0.2 T\alpha_c^2.   \hspace{10cm}     (17)
\nonumber
\ee
We could impose on this equation the initial condition $r(T_f) \approx R_{eq} (T_f)$, in analogy with
what we did before in calculating the persistent burning-out of residual thetonium. It turns out, however,
that the answer does  not depend on the initial conditions (in analogy with what happens in the case of 
residual magnetic monopoles [8]). Some simple steps lead to the result
\be
r(q_\theta) \approx \frac{10 T}{\alpha_c^2 m_P} \approx 10^{-18} \frac{T}{\Lambda_c},
  \hspace{8cm}     (18)
\nonumber
\ee
where $\Lambda_c \leq T$, For $T < \Lambda_c$ there must be no appreciable burnout of the quarks,
since they adhere to ordinary hadrons and their kinetics is determined by the strong hadron interactions.
The result (18) raises a serious difficulty for the hypothesis that $\theta$ quarks exist, since the
observed restrictions are much lower. A similar conclusion was reached in Ref. [5].\\

One possibility for avoiding this difficulty is to assume that $q_\theta$ is unstable; these quarks could 
decay, for example, via the channel
\be
q_\theta \rar q + l_\theta + ....  
\nonumber
\ee 
with conservation of the color and $\theta $ charges. If the $\theta$ quarks are fermions, then either
$l_\theta$ must be a boson, or else another fermion must be emitted in this decay ($q,\,\nu\,,3q$ etc.).
The specific form of the interaction depends upon the quantum numbers of $q_\theta$.\\

This it appears that the existence of stable $\theta$ quarks is impossible, but  unstable ones that decay
into $\theta$ leptons can exist.

As for electrically charged $\theta$ leptons, the hypothesis that they exist encounters the same difficulties
as that of $\theta$ quarks, and therefore they must decay into neutral $\theta$ leptons [5]. The latter can
exist in large numbers without contradicting the observations, and furthermore they could contribute to
solving the problem of the missing mass in the Universe.  We note also that the restriction (15) may not
hold even if there is an asymmetry in $L_\theta$, if $theta$-singlet states composed of $\theta$ leptons
can decay into ordinary particles  with violation of $L_\theta$ conservation (in analogy with the possibility
of proton decay). The point is that with increase of the mass of a $\theta$ lepton the probability of decay
with nonconservation of  $L$ increases$^2)$:
\be
\Gamma (\Delta L_\theta \neq 0) \sim 
\frac{\alpha^2}{m^4_X}\,m_l^5\,\left(\frac{\Lambda_\theta}{m_l}\right)^3,
\hspace{8cm}     (19)
\nonumber
\ee
where $m_X$ is the mass of the intermediate boson which accomplishes the great synthesis, 
$m_X \approx 10^{15} $ GeV, and $\alpha \approx 10^{-2}$ is the unique coupling constant; the 
factor $(\Lambda_\theta /m_l)^3$ is associated with the large size of the bound state constructed from
neutral $\theta$ leptons. Requiring that  $\Gamma^{-1}$ be smaller than the lifetime of the Universe, 
$t_U = 10^{17}$ sec, we find  that if $m_l \gtrsim 100 m_N$, then
\be
m_l > 3\cdot 10^{11} m_N (m_N /\Lambda_\theta )^{3/2}.
\hspace{8cm}     (20)
\nonumber
\ee
Actually the restriction is even stronger, since  the decay, recent in the cosmological scale, of large 
numbers of heavy particles constructed from $l_\theta^0$ would lead to a contradiction with the observe
 spectrum of cosmic rays.\\
 
 Let us go back to the restriction (14) on the mass of thetonium. In its derivation it was tacitly assumed 
 that thetonium is practically stable. According to Ref. [5] the lifetime of thetonium for decay into two
 photons is given by
 \be
 \tau = \alpha_s^{-2} m_N^{-1} \left(\frac{m_\theta}{\Lambda_\theta}\right)^{9} \frac{ m_N}{m\theta},
\nonumber
\ee
where $m_\theta$ is the mass of charged $\theta$ particles.\\

If $\tau < 10^5 $ sec, the photons produced in the decay are thermalized in the primary plasma and 
cause no conflict with the present observations. This leads to an allowed region in values of 
$\Lambda_\theta$ and $m_\theta$:
\be
\Lambda_\theta > 10^{-3} m_\theta  \left( 100 m_N / m_\theta \right)^{1/9}
\nonumber
\ee
Since we know that $m_\theta > 15 $ GeV (otherwise charged $\theta$ fermions would have been 
observed in$e^+e^-$ collisions), we have $\Lambda_\theta > 20$ MeV; this is an interestng regkon,
since from the beginning we have been concerned with particles with of confinement radius  much
smaller than that of ordinary hadrons.\\

The range of lifetimes $10^5$ sec $<\tau_\theta < 10^{17} $ sec is excluded by the observed data on
the spectrum of the cosmic electromagnetic radiation, so that $\tau_\theta$ must be larger that the age
of the Universe  $t_U = 10^{17}$ sec, and accordingly the restriction
\be 
\Lambda_\theta < 3\cdot 10^{-5} m_\theta\, \left(100 m_N /m_\theta \right)^{1/9}
 \hspace{8cm}     (21)
\nonumber
\ee
must hold. If we suppose that the masses of all $\theta$ fermions are of the same order of magnitude,
than from the  condition (15), which strictly speaking is valid for neutral $\theta$ leptons, we get 
$\Lambda_\theta < 3$ MeV. \\

The work was stimulated by many discussions with L.B. Okun', to whom  am deeply grateful. \\[4mm]

$^1)$ The subscript $f$ here refers to the "freezing" of the concentration.\\
$^2)$ A similar situation was pointed out by Ya. B. Zel'dovich for the hypothesis of a stable heavy
lepton.\\

$--------------------------- $\\
1. Ya. B. Zel'dovich and I.D. Novikov, Stroenie i evolyutsiya Vselennoi (Structure and evoluiton of the 
Universe), Nauka, 1975. \\
2. A.D. Dolgov and Ya.B. Zel'dovich,  Usp. Fiz. Nauk {\bf 130}, 559 (1990) 
[Rev. Mod. Phys. {\bf 53}, 1 (1981)].\\
3. Ya.B.  Zel'dovich, L.B. Okun', and S.B.  Pikel'ner, Usp. Fiz. Nauk, {\bf 87}, 113 (1965) 
[Sov. Phys. Uspekhi {\bf 8}, 702 (1966)]. \\
4. Ya.B. Zel'dovich, Zh. Eksp. Teor. Fiz. {\bf 48}, 986 (1965) [Sov. Phys. JETP {\bf 21}, 656 (1965)].\\
5. L.B.Okun',  Pis'ma Zh. Eksp. Teor. Fiz. {\bf 31}, 156 (1980) 
[JETP Lett.  {\bf 31}, 144 (1980]; Preprint No. 6,
Int. Theor. Exp. Phys. Moscow.\\
6. S.S. Gershtein and Ya.B. Zel'dovich,  Pis'ma Zh. Eksp. Teor. Fiz. {\bf 4}, 174 (1966) 
[JETP Lett.  {\bf 4}, 120 (1966)].\\
7. A.V. Novikov, M.A. Shifman, A.I. Vainshteiin, and V.I. Zakharov, Phys. Lett. {\bf 66B} 347 (1979).\\
8. Ya.B. Zel'dovich and M.Yu. Khlopov, Phys. Lett. {\bf 79 B}, 239 (1978).\\[3mm]
Translated by W.H. Furry.

\begin{figure}[h]
\hskip -14mm
\includegraphics{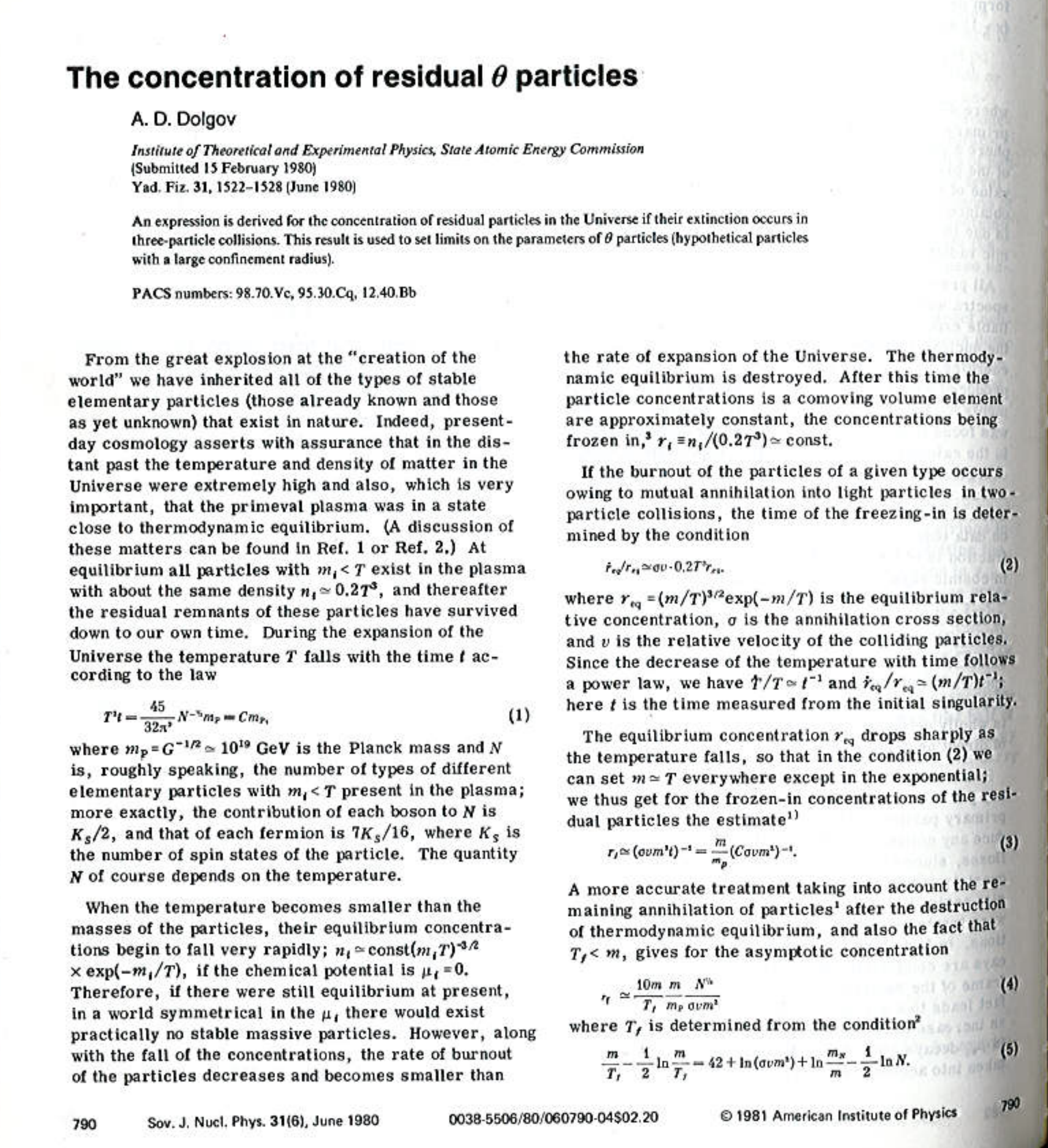}
\end{figure}

\end{document}